\documentclass[12pt]{article}
\usepackage{epsf}
\def\be{\begin{equation}}
\def\ee{\end{equation}}
\def\bea{\begin{eqnarray}}
\def\eea{\end{eqnarray}}

\begin{document}
\begin{titlepage}
\begin{center}
{\Large \bf William I. Fine Theoretical Physics Institute \\
University of Minnesota \\}
\end{center}
\vspace{0.2in}
\begin{flushright}
FTPI-MINN-18/03 \\
UMN-TH-3711/18 \\
February 2018 \\
\end{flushright}
\vspace{0.3in}
\begin{center}
{\Large \bf $S$ wave hidden charm - hidden strangeness production in $e^+e^-$ annihilation 
\\}
\vspace{0.2in}
{\bf  M.B. Voloshin  \\ }
William I. Fine Theoretical Physics Institute, University of
Minnesota,\\ Minneapolis, MN 55455, USA \\
School of Physics and Astronomy, University of Minnesota, Minneapolis, MN 55455, USA \\ and \\
Institute of Theoretical and Experimental Physics, Moscow, 117218, Russia
\\[0.2in]

\end{center}

\vspace{0.2in}

\begin{abstract}
It is suggested that the recently observed  enhancement of the relative yield in $e^+e^-$ annihilation of states with hidden charm and hidden strangeness  above 4.43\,GeV is associated with the $S$ wave production of the charmed strange meson pairs $D_{s0}(2317) \bar D_s^* + {\rm c.c.}$ and $D_{s1}^*(2460) \bar D_s+ {\rm c.c.}$. This mechanism implies a  pattern of breaking of the Heavy Quark Spin Symmetry (HQSS) that can be tested in the final channels such as $\eta_c(1S) \phi$, $h_c(1P) \eta $, $h_c(1P) \eta' $ produced in the same energy range. 
\end{abstract}
\end{titlepage}

The $e^+e^-$ annihilation at the center of mass (c.m.) energy above the threshold for open charm is well known to exhibit a complex pattern of production of pairs of charmed mesons as well as final states with charmonium and light mesons. In particular, some final channels with charmonium and light mesons reveal peaks in the energy behavior, such as  $Y(4260)$ in the yield of $\pi \pi J/\psi$, or $Y(4360)$ in the $\pi \pi \psi(2S)$ channel~\cite{pdg}. The properties of these peaks are not yet fully established, and  the annihilation cross section to other channels (e.g.  $\pi \pi h_c$~\cite{beshc}, or $\omega \chi_{c0}$~\cite{besomega}) displays peaks around the same energy with somewhat different parameters, and it is also quite likely that at least some of the peaks emerge from a multiquark or heavy meson-antimeson dynamics (a discussion and further references can be found e.g. in Ref.~\cite{krs}).
A recent experimental study~\cite{beskk} of the production of the final channel $K \bar K J/\psi$ in the c.m. energy range 4.19 -- 4.6 GeV indicates a significant enhancement of the ratio $\sigma(K \bar K J/\psi)/\sigma (\pi \pi J/\psi)$ above 4.43\,GeV, which enhancement is referred to in Ref.~\cite{beskk} as `evidence for a structure around 4.5\,GeV'. Such structure, if confirmed, may (in principle) be related to the peaks in channels without hidden strangeness in the context of exotic four-quark resonances, albeit at a higher energy due to the extra mass of the strange quark and antiquark. However the so far available data do not necessarily require such interpretation and 
the purpose of the present letter is to explore a simpler description of the observed enhancement, not associated with new resonances in $e^+e^-$ annihilation, but rather related to an opening hidden charm - hidden strangeness continuum production of the pairs of charmed strange mesons of the type $D_{s0}(2317) \bar D_s^*+ {\rm c.c.}$ and $D_{s1}^*(2460) \bar D_s + {c.c.}$, of which final channels the first one has in fact been observed~\cite{bes}. The peculiarity of these meson pairs is that unlike most of  other heavy meson states they can be produced in the $e^+e^-$ annihilation in the $S$-wave resulting in a significant growth of the yield with energy right above the threshold. Furthermore, the excited $D_{s0}(2317)$ and $D_{s1}^*(2460)$ charmed strange mesons are narrow, and the threshold for their production is well defined, unlike for their non-strange counterparts, supposedly $D_0^*(2400)$ and $D_1(2430)$~\cite{pdg}, that have large widths in the range 200 - 400\,MeV. As will be discussed, in the threshold region for these channels and up until the threshold for the $D_{s1}^*(2460) \bar D^*_s + {c.c.}$ at about 4.57\,GeV a particular spin structure for the heavy $c \bar c$ quark pair is to be expected, which can be translated into specific expectations for the yield in channels with spin-singlet states of charmonium  and light mesons with hidden strangeness. In addition it will be argued that the considered meson pair with hidden strangeness should strongly re-scatter into similar meson pairs without the strange (anti)quarks containing the very broad excited charmed mesons. For this reason it is unlikely that any features exist in the energy behavior of the cross section for the strange - antistrange meson pairs production with a scale smaller than the widths of the broad non-strange excited charmed mesons.

As is well known, the non-strange scalar and axial charmed mesons are broad due to the very strong $S$ wave decay: $D_0^*(2400) \to D \pi$ and $D_1(2430) \to D^* \pi$. For their  charmed strange counterparts similar decays with emission of a Kaon are impossible kinematically, so that their hadronic decays proceed with violation of isospin: $D_{s0}(2317) \to D_s \pi^0$ and $D_{s1}(2460) \to D^*_s \pi^0$, which accounts for  very narrow (and currently unmeasured) widths of these mesons. The kinematical consideration however does not apply to a rescattering with destruction of hidden strangeness: $D_{s0}(2317) \bar D_s^* \to D \bar D_1(2430)$ and  $D_{s1}^*(2460) \bar D_s \to D^* \bar D^*_0(2400)$ through the Kaon exchange, as shown in Fig.~1. The light pseudoscalar interaction vertex in this process is the same as in the strong decays $D_0^*(2400) \to D \pi$ and $D_1(2430) \to D^* \pi$ due to the flavor $SU(3)$
symmetry, so that the amplitude of production of the pairs of strange mesons is likely to be dominated by the absorptive part due to rescattering into meson pairs without hidden strangeness. Then the broad width of the excited non-strange charmed mesons generally results in a smearing of any features in the hidden strangeness channel over energy range comparable with the widths of the non-strange mesons. For instance there should be no (relatively) narrow `molecular' resonances made from a combination of $D_{s0}(2317) \bar D_s^*$ and  $D_{s1}^*(2460) \bar D_s$.  In particular, this behavior makes unlikely a significant admixture of the pairs $D_{s0}(2317) \bar D_s^*+ {\rm c.c.}$ and $D_{s1}^*(2460) \bar D_s + {c.c.}$ in the $\psi(4415)$ resonance as discussed in the literature~\cite{cz}. 

\begin{figure}[ht]
\begin{center}
 \leavevmode
    \epsfxsize=8cm
    \epsfbox{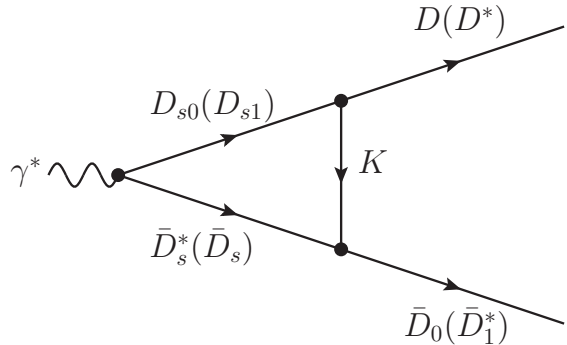}
    \caption{Rescattering with destruction of hidden strangeness. All vertices in this graph are $S$-wave.}
\end{center}
\end{figure} 

The energy range where the effect of the discussed here hidden charm - hidden strangeness meson pairs can be practically studied extends from the threshold at 4.43\, GeV to the maximal available energy in the BESIII experiment 4.6\,GeV. Due to the expected smooth behavior the production amplitude can be approximated in this range by a constant $A$. Furthermore, it has been recently argued~\cite{mv18} that the amplitude should be the same in the channels $D_{s0}(2317) \bar D_s^*+ {\rm c.c.}$ and $D_{s1}^*(2460) \bar D_s + {c.c.}$: 
\be
A[e^+e^- \to D_{s0}(2317) \bar D_s^*] = A[e^+e^- \to D_s \bar D_{s1} (2460) ] = A~,
\label{eqa}
\ee
which equality is protected within the chiral symmetry scheme~\cite{beh} as well as by the HQSS~\cite{mvbsds}. The cross section in both channels should then (approximately) follow the excitation behavior for $S$ wave production:
\be
\sigma [e^+e^- \to D_{s0}(2317) \bar D_s^*+ {\rm c.c.}] = \sigma [e^+e^- \to D_s \bar D_{s1} (2460) + {\rm c.c.}] \propto \sqrt{E_{cm}-E_0}~,
\label{eqs}
\ee
with $E_{cm}= \sqrt{s}$ being the c.m. energy and $E_0 \approx 4.43$\,GeV is the threshold energy practically coinciding for both channels. Clearly the behavior described by Eq.(\ref{eqs}) can be tested by the BESIII experiment. In fact the process $e^+e^- \to D_{s0}(2317) \bar D_s^*+ {\rm c.c.}$ has been observed~\cite{bes}. However no result is reported yet on energy dependence of the cross section, neither on the production of the other channel $D_s \bar D_{s1} (2460) + {\rm c.c.}$.

The composition of the produced two-meson $S$-wave state in terms of the total spin  $S_H^{PC}$ of the $c \bar c$ quark pair and the total spin state $S_L^{PC}$ of the light degrees of freedom is found as~\cite{mvbsds}
\be
{1 \over 2} \left ( D_s^* \bar D_{s0} - \bar D^*_s D_{s0} + D_{s1} \bar D_s - \bar D_{s1} D_s \right ) :~~ {1 \over \sqrt{2}} \, 1^{--}_H \otimes 0^{++}_L + {1 \over \sqrt{2}} \, 0^{-+}_H \otimes 1^{+-}_L~,
\label{spind}
\ee
and describing a presence of an ortho- spin state $1^{--}_H$ of the heavy $c \bar c$ pair as well as a para- state $0^{-+}_H$. For this reason a significant HQSS-violating presence of  spin-singlet charmonium states  should be expected in the final channels with charmonium and light mesons that are produced due to a rescattering of the discussed meson pairs as shown in Fig.~2. This behavior should persist up to the heavier threshold for the $D_{s1}(2460) \bar D^*_s$ at $E_1 \approx 4.57\,$GeV. The $S$-wave state of the vector and axial mesons has the following  heavy light spin decomposition~\cite{mvbsds}
\be
\left . \left (  D^*_s \bar D_{s1}  - \bar D^*_s D_{s1}  \right ) \right |_{J^{PC}=1^{--}} :~~  {1 \over \sqrt{2}} \, 1^{--}_H \otimes 0^{++}_L - {1 \over \sqrt{2}} \,  0^{-+}_H \otimes 1^{+-}_L~~. 
\label{spind1}
\ee
The production amplitude near the threshold is related to that for the lighter state [Eq.(\ref{eqa})] as 
\be
A[e^+e^- \to D^*_s \bar D_{s1} (2460) ] = A~,
\label{eqa1}
\ee
which relation ensures restoration of HQSS at energy sufficiently above the higher threshold due to the cancellation of the $ 0^{-+}_H \otimes 1^{+-}_L$ spin component from the expressions (\ref{spind}) and (\ref{spind1}).  

\begin{figure}[ht]
\begin{center}
 \leavevmode
    \epsfxsize=8cm
    \epsfbox{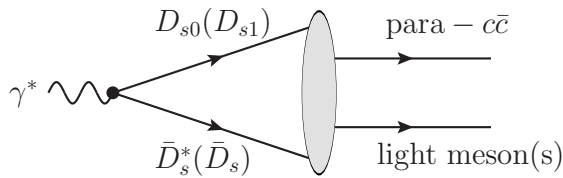}
    \caption{Production of HQSS-violating para-charmonium and light mesons through heavy mesons with hidden strangeness. The grayed ellipse corresponds to the conversion of heavy meson pair to charmonium plus light meson(s), which amplitude is unknown, but is assumed to conserve the heavy quark spin.}
\end{center}
\end{figure}

One can then evaluate the energy dependence of the yield of a spin singlet charmonium state due to mechanism of Fig.~2 in the approximation of a constant production amplitude. In this approximation the constant parts of the loop near the heavy meson threshold cancel between the contribution of the ligter channels $D_{s0}(2317) \bar D_s^*+ {\rm c.c.}$ and $D_{s1}^*(2460) \bar D_s + {c.c.}$ and that of the heavier state  $D^*_s \bar D_{s1} (2460) + c.c.$ with only the energy - dependent absorptive part (due to unitarity)  non vanishing in the threshold region:
\be
A(e^+e^- \to {\rm para-}c \bar c + X) \propto [\sqrt{E_{cm}-E_0} \, \theta(E_{cm}-E_0) - \sqrt{E_{cm}-E_1} \, \theta(E_{cm}-E_1)]
\label{absp}
\ee
with $\theta$ being the step function. The  HQSS-violating channels with para-charmonium and hidden strangeness in the final state, to which the formula (\ref{absp}) can be applied are $h_c \eta$, $h_c \eta'$ and $\eta_c \phi$. The expected behavior for the yield of these channels corresponding to Eq.(\ref{absp}) should have a cusp at the threshold $E_1$ as  is shown in Fig.~3. One of these processes, $e^+e^- \to h_c \eta$ has been actually observed~\cite{besetac}, with an indication that the cross section has a `bump' between 4.4 and 4.6\,GeV. It can be noted that the observed production of this final state at lower energies may proceed through other mechanisms of HQSS breaking (such as discussed e.g. in Ref.~\cite{lv})  not associated with hidden strangeness. The contribution of such mechanisms should die away above the thresholds for the states without hidden strangeness, so that it is likely that the discussed here mechanism for production dominates the (apparent) HQSS symmetry breaking at the energy aroung 4.5\,GeV. The relative composition of the contributing states with and without hidden strangeness can be tested by the relative yield of $h_c \eta$ and $h_c \eta'$.  It is clear however that the channel $\eta_c \phi$ appears to be significantly more pure with regards to the contribution of states with hidden strangeness, and can thus be more indicative of such contribution.

\begin{figure}[ht]
\begin{center}
 \leavevmode
    \epsfxsize=8cm
    \epsfbox{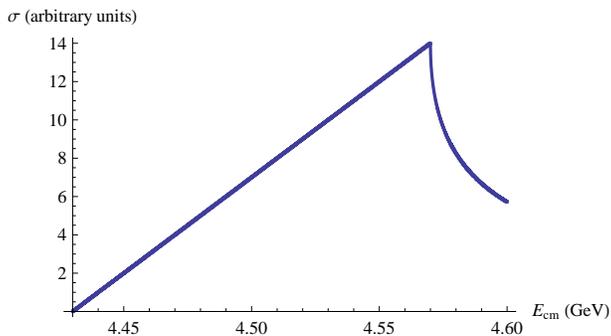}
    \caption{The energy behavior of the cross section for HQSS breaking final states ($h_c \eta$, $\eta_c \phi$).}
\end{center}
\end{figure}

The channels with ortho-charmonium in the final state are HQSS conserving. In particular, the contribution of the lighter and the heavier meson pairs to the production amplitude adds constructively, so that the amplitude near the thresholds can be generally written as 
\be
A(e^+e^- \to {\rm ortho-}c \bar c + X) = B + i C \, \left  [\sqrt{E_{cm}-E_0} \, \theta(E_{cm}-E_0) + \sqrt{E_{cm}-E_1} \, \theta(E_{cm}-E_1) \right ]~,
\label{aort}
\ee
where $B$ and $C$ are the coefficients (with the same complex phase) corresponding to the dispersive and the absorptive part of the meson loop. Clearly, the relative rise of the yield above the threshold at $E_0$ is determined by the ratio $|C/B|^2$. Therefore within the discussed picture such rise suggested by the experiment~\cite{beskk} can be interpreted in terms of a sizable value of this ratio. It however appears impossible to quantitatively determine this value, given the uncertainty in the present data. Another type of final states with ortho-charmonium and hidden strangeness are $\chi_{cJ} \phi$ of which the production of $\chi_{c1} \phi$ and $\chi_{c2} \phi$ has been observed~\cite{beschiphi} at the c.m. energy 4.6\,GeV. However the existing data appear too scarce for a meaningful quantitative analysis. 

In summary. It is suggested that the observed rise in the yield in $e^+e^-$ annihilation of hidden charm - hidden strangeness channels at c.m. energy above approximately 4.43\,GeV is associated with $S$ wave meson pairs $D_{s0}(2317) \bar D_s^*+ {\rm c.c.}$ and $D_{s1}^*(2460) \bar D_s + {c.c.}$. The cross section for these pairs is expected to be smooth with energy due to their strong rescattering into very broad non-strange excited charmed mesons. The discussed mechanism suggest a distinctive pattern of HQSS violation and can be tested by the yield in the energy range 4.43 - 4.6\,GeV of final states with para-charmonium and light mesons with hidden strangeness, such as $\eta_c \phi$, $h_c \eta$ and $h_c \eta'$.

This work is supported in part by U.S. Department of Energy Grant No.\ DE-SC0011842.

\end{document}